# A Connection Between Gravitation and Electromagnetism

Douglas M. Snyder
Los Angeles, California[1]

   It is argued that there is a connection between the fundamental forces of electromagnetism and gravitation. This connection occurs because of: 1) the fundamental significance of the finite and invariant velocity of light in inertial reference frames in the special theory, and 2) the reliance of the general theory of relativity upon the special theory of relativity locally in spacetime.[2] The connection between the fundamental forces of electromagnetism and gravitation follows immediately from these two points. Because this connection has not been acknowledged, a brief review is provided of: 1) the role of the finite and invariant velocity of light in inertial reference frames in the special theory, and 2) certain fundamental concepts of the general theory, including its reliance on the special theory locally.

## 1.  *The Velocity of Light in the Special Theory*

   It is known that the reason that the Lorentz transformation equations of special relativity differ from the Galilean transformation equations underlying Newtonian mechanics is due to the finite value for the invariant velocity of light in any inertial reference frame (i.e., a spatial coordinate system attached to a physical body and for which Newton's first law of motion holds) that is used to develop simultaneity and time in general in any inertial reference frame. Other results obtained in the special theory of relativity depend on the Lorentz transformation equations, as opposed to the Galilean transformation equations, and are thus dependent on the finite and invariant velocity of light.[1,2,3]

   For an inertial reference frame $W'$ with one spatial dimension, $x'$, in uniform translational motion relative to an inertial reference frame $W$ (with one spatial dimension, $x$) with velocity $v$ along the $x$ and $x'$ axes in the direction of increasing values of $x$ and $x'$, the Lorentz transformation equations may be stated as:

$$x' = (x - vt)/(1 - v^2/c^2)^{1/2} \tag{1}$$

and

$$t' = [t - (v/c^2)x]/(1 - v^2/c^2)^{1/2} \ . \tag{2}$$

# Gravitation and Electromagnetism

*c* is the finite and invariant value for the velocity of light in inertial reference frames. *c* is found in equations 1 and 2 because of the fundamental role of the velocity of light in the development of simultaneity and time in general in inertial reference frames in the special theory.[1] If an existent with an arbitrarily great velocity (also represented by *c*) were used instead of light to develop simultaneity in an inertial reference frame, *c* would be arbitrarily large and equations 1 and 2 would become:

$$x' = x - vt \tag{3}$$

and

$$t' = t \tag{4}$$

Equations 3 and 4 are Galilean transformation equations for W and W' for the circumstances described.

As an example of the significance of the finite and invariant velocity of light in the special theory, it can be noted that the Lorentz transformation equations underlie the derivation of the force law in the special theory as well as the transformation of force in inertial reference frames in uniform translational motion relative to one another.[4,5] These transformation of force equations in the special theory allow for the invariance of Maxwell's equations of electromagnetism in inertial reference frames in uniform translational motion relative to one another.[6] In fact, one can use the Lorentz transformation equations to transform the field differentials in Maxwell's laws of electromagnetism for two inertial reference frames in uniform translational motion relative to one another while holding that these laws should be invariant. Doing this, one finds that the transformations for the electric and magnetic components of the field indicated are exactly those that one finds by application of the transformation equations for force in the special theory to the electric and magnetic components of the electromagnetic field.

It should be remembered that one of Maxwell's most fundamental results was the identification of electromagnetic radiation with light because the measured velocity of light in vacuum is in accord with the velocity of electromagnetic radiation in vacuum derived using Maxwell's equations. It should be emphasized that this identification of light with electromagnetic radiation through Maxwell's equations provides the theoretical foundation for the empirical evidence that the velocity of light in vacuum is finite and invariant and thus does not depend on the velocity of the emitting body relative to the observer at rest in an inertial reference frame.

# Gravitation and Electromagnetism

## 2. *The Dependence of the General Theory on the Special Theory*

### 2.1 The Principle of Equivalence

It is known that in the general theory of relativity gravitation corresponds to spacetime curvature. Spacetime curvature may be developed by considering an accelerating reference frame or pattern of such reference frames as a pattern of local Lorentz frames moving at slightly different uniform, translational velocities relative to one another. The accelerating reference frame, or frames, in general corresponds to an inertial reference frame experiencing a gravitational field because the descriptions of physical phenomena in these circumstances are equivalent. In particular, the descriptions of the motion of physical bodies when described from a uniformly accelerating reference frame or an inertial reference frame experiencing a gravitational field of uniform intensity are equivalent.[7,8,9]

Thus, a uniformly accelerating reference frame serves as a bridge between the gravitation-free inertial reference frame from which the motion of the uniformly accelerating reference frame is judged and the inertial reference frame experiencing a gravitational field of uniform intensity that is no different for the description of physical phenomena to the uniformly accelerating reference frame. The inertial reference frame experiencing the gravitational field of uniform intensity is equivalent to the gravitation-free inertial reference frame as concerns the description of physical phenomena. The principle of equivalence provides the theoretical foundation for the basic equality of inertial and gravitational mass and allows for the in general equivalence between accelerating reference frames and inertial reference frames experiencing a gravitational field.[9,10,11]

In *The Meaning of Relativity*, Einstein stated the principle of equivalence in the following way and indicated how it provides the theoretical underpinning for the essential equality of inertial and gravitational mass:

> Let now $K$ be an inertial system. Masses which are sufficiently far from each other and from other bodies are then, with respect to $K$, free from acceleration. We shall also refer these masses to a system of co-ordinates $K'$, uniformly accelerated with respect to $K$. Relatively to $K'$ all the masses have equal and parallel accelerations; with respect to $K'$ they behave just as if a gravitational field were present and $K'$ were unaccelerated. Overlooking for the present the question as to the "cause" of

> such a gravitational field...there is nothing to prevent our conceiving this gravitational field as real, that is, the conception that $K'$ is "at rest" and a gravitational field is present we may consider as equivalent to the conception that only $K$ is an "allowable" system of co-ordinates and no gravitational field is present. The assumption of the complete physical equivalence of the systems of coordinates, $K$ and $K'$, we call the "principle of equivalence;" this principle is evidently intimately connected with the law of the equality between the inert and the gravitational mass, and signifies an extension of the principle of relativity to co-ordinate systems which are in non-uniform motion relatively to each other. In fact, through this conception we arrive at the unity of the nature of inertia and gravitation. For according to our way of looking at it, the same masses may appear to be either under the action of inertia alone (with respect to $K$) or under the combined action of inertia and gravitation (with respect to $K'$).[12]

*2.2    The Special Theory Holds Locally*

As noted, the derivation of the spacetime curvature associated with an accelerating reference frame may be based on the notion that an accelerating reference frame is composed of a pattern of local Lorentz frames (essentially, local, inertial reference frames). These local Lorentz frames are considered to have varying uniform, translational velocities, and the Lorentz transformation equations of special relativity thus hold for these local reference frames.

In outlining the essentials of the general theory of relativity, Einstein discussed an example of how an observer in an accelerating reference frame would be affected in his measurements of temporal duration and how his or her measurements could be accounted for in terms of a local Lorentz frame. He wrote:

> Let us consider a space-time domain in which no gravitational field exists relative to a reference-body $K$ whose state of motion has been suitably chosen. $K$ is then a Galileian reference-body as regards the domain considered, and the results of the special theory of relativity hold relative to $K$. Let us suppose the same domain referred to a second body of reference $K'$, which is rotating uniformly with respect to $K$. In order to fix our ideas, we shall imagine $K'$ to be in the form of a plane circular disc, which rotates uniformly in its own plane about its centre. An observer who is sitting eccentrically on the disc $K'$ is sensible of a force which acts outwards in a radial direction, and which would be interpreted as an effect of inertial (centrifugal force) by an observer who was at rest with respect to the original reference-body $K$. But the observer on the disc may regard his disc as a reference-body which is "at rest"; on the basis of the general principle of relativity he is justified in doing this. This force acting on himself, and in fact on all other bodies which are at rest relative to the disc, he regards as the effect of a gravitational field. But since the observer believes in the general theory of relativity, this does not disturb him: he is quite in the right when he believes that a general law of gravitation can be formulated–a law which not only explains the motion of the stars correctly, but also the field of force experienced by himself.

The observer performs experiments on his circular disc with clocks and measuring-rods. In doing so, it is his intention to arrive at exact definitions for the signification of time- and space-data with reference to the circular disc $K'$, these definitions being based on his observations. What will be his experience in this enterprise?

To start with, he places one of two identically constructed clocks at the centre of the circular disc, and the other on the edge of the disc, so that they are at rest relative to it. We now ask ourselves whether both clocks go at the same rate from the standpoint of the non-rotating Galileian reference-body $K$. As judged from this body, the clock at the centre of the disc has no velocity, whereas the clock at the edge of the disc is in motion relative to $K$ in consequence of the rotation....the latter clock goes at a rate permanently slower than that of the clock at the centre of the circular disc, i.e., as observed from $K$. It is obvious that the same effect would be noted by an observer whom we will imagine sitting alongside his clock at the centre of the circular disc. Thus on our circular disc, or, to make the case more general, in every gravitational field, a clock will go more quickly or less quickly, according to the position in which the clock is situated (at rest).[13]

What is the general relation for the durations of some occurrence in two inertial reference frames moving in a uniform translational manner relative to one another? Consider the inertial reference frames W and W' discussed earlier. If this occurrence maintains its position in W', then the relation between the durations of this occurrence in W and W' is given by:

$$\Delta t = \Delta t'/(1 - v^2/c^2)^{1/2} \ . \tag{5}$$

Spacetime curvature may then be found for an accelerating frame, or an associated inertial reference frame in a gravitational field, essentially by deriving the pattern of special relativistic results that hold for the local Lorentz frames and that compose the global reference frame.[14,15,16] Specifically, the curvature of the spacetime continuum can be developed through the use of a field tensor, $g_{ik}$. The use of this tensor allows for local Lorentz frames while also allowing that these local Lorentz frames do not have to be identical in terms of their spacetime characteristics. Thus, there is the possibility of a pattern of different, local Lorentz frames, the result of which is a curved spacetime continuum. The particular coordinate scheme that one applies to the spacetime continuum is arbitrary as long as it is Gaussian in nature. $g_{ik}$ are certain functions of the Gaussian coordinate scheme that transform for a continuous coordinate transformation. $g_{ik}$ has a symmetrical property (i.e., $g_{ik} = g_{ki}$) that holds in general for a Gaussian coordinate scheme applied to the spacetime continuum. $g_{ik}$ must be able to account for global, gravitation-free inertial reference frames. In the case of such inertial reference frames, $g_{ik}$ reduces to a particular form such that the resulting metric is that characteristic of a Lorentz reference frame. The general covariant field law that applies to $g_{ik}$ of different representations of the spacetime continuum is given by a set of differential equations in the form of another symmetrical tensor, $R_{ik}$.[17,18]

Misner, Thorne, and Wheeler summed up the relation of the general and special theories in writing, "General relativity is built on special relativity".[19] In elaborating on this statement, the authors wrote:

> A tourist in a powered interplanetary rocket feels "gravity." Can a physicist by local effects convince him that this "gravity" is bogus? Never, says Einstein's principle of the local equivalence of gravity and accelerations. But then the physicist will make no errors if he deludes himself into treating true gravity as a local illusion caused by acceleration. Under this delusion, he barges ahead and solves gravitational problems by using special relativity: if he is clever enough to divide every problem into a network of local questions, each solvable under such a delusion, then he can work out all influences of any gravitational field. Only three basic principles are invoked: special relativity physics, the equivalence principle, and the local nature of physics. They are simple and clear. To apply them, however, imposes a double task: (1) take spacetime apart into locally flat pieces (where the principles are valid), and (2) put these pieces together again into a comprehensible picture. To undertake this dissection and reconstitution, to see curved dynamic spacetime inescapably take form, and to see the consequences for physics: that is general relativity.[19]

### 3.   *The General Principle of Relativity*

The principle of equivalence led Einstein to maintain that all frames of reference should be equivalent for the expression of physical law. The conceptual basis of space-time structure in the general theory has been noted and has been shown to be dependent upon its description locally using the special theory. With these aspects of the general theory in mind, the general principle of relativity can be more precisely stated as:

> All Gaussian co-ordinate systems are essentially equivalent for the formulation of the general laws of nature.[20]

Einstein wrote in his original paper proposing the general theory:

> In the general theory of relativity, space and time cannot be defined in such a way that differences of the spatial co-ordinates can be directly measured by the unit measuring-rod, or differences in the time co-ordinate by a standard clock.
>
> The method hitherto employed [in the special theory as well as Newtonian mechanics adhering to Euclidean geometry] for laying co-ordinates into the space-time continuum in a definite manner thus breaks down, and there seems to be no other way which would allow us to adapt systems of co-ordinates to the four-dimensional universe so that we might expect from their application a particularly simple formulation of the laws of

Gravitation and Electromagnetism

nature. So there is nothing for it but to regard all imaginable systems of co-ordinates, on principle, as equally suitable for the description of nature. This comes to requiring that:–

*The general laws of nature are to be expressed by equations which hold good for all systems of coordinates, that is, are covariant with respect to any substitutions whatever (generally covariant).*[21]

The use of arbitrary Gaussian coordinate schemes, and their accompanying metrical coefficients $g_{ik}$, to represent the spacetime continuum allows for the development of the general formulation of physical law that applies to these coordinate schemes. As noted, $g_{ik}$ transform for a continuous coordinate transformation. $R_{ik}$ is a formulation of gravitational field law for $g_{ik}$.

*4. Light and the General Theory*

An important result of relativity theory, specifically the special theory, is that light has properties associated with mass. This is a consequence of the general equivalence of mass and energy in the special theory. The general equation relating mass and energy is

$$E = mc^2 , \qquad (6)$$

where $E$ is the energy equivalent of the relativistic mass of a physical entity, $m$ is this relativistic mass, and $c$ is the finite and invariant velocity of light in inertial reference frames. In the special theory, the kinetic energy of a physical existent can be expressed by the equation

$$K = mc^2 - m_0 c^2 , \qquad (7)$$

where $K$ is the kinetic energy, $m$ is the relativistic mass of the physical existent, $m_0$ is the rest mass of this existent, and $c$ is the finite and invariant velocity of light in inertial reference frames. Another form for the kinetic energy is

$$K = m_0 c^2 [1/(1 - v^2/c^2)^{1/2} - 1] , \qquad (8)$$

where $v$ is the uniform, translational velocity of the inertial reference frames relative to one another. The relation between the relativistic mass of a physical existent, $m$, and its rest mass, $m_0$, is given by

$$m = m_0/(1 - v^2/c^2)^{1/2} , \qquad (9)$$

and the momentum is given by

Gravitation and Electromagnetism

$$p = [m_0/(1 - v^2/c^2)^{1/2}]v \quad . \tag{10}$$

Combining equations 8 and 10 by removing *v* results in

$$(K + m_0 c^2)^2 = (pc)^2 + (m_0 c^2)^2 \quad . \tag{11}$$

Given

$$E = K + m_0 c^2 \quad , \tag{12}$$

$$E^2 = (pc)^2 + (m_0 c^2)^2 \quad .^{(22)} \tag{13}$$

Light has no rest mass in the special theory as it is never at rest according to the empirically verified postulate of the special theory regarding the invariant and finite velocity of light in inertial reference frames. But light does have momentum, *p*. For light,

$$E = pc \tag{14}$$

in the special theory, a relation found in classical electromagnetic theory. Light therefore has effective mass.

In sensitive enough conditions, the influence of a gravitational field on light should thus be detectable. Using the general theory, Einstein predicted that a light ray would be deflected toward the sun in accordance with the following equation

$$\mu = 1.7 \text{ sec. of arc}/\Delta \tag{15}$$

where $\mu$ is the angle of deflection and $\Delta$ is the distance of the light ray from the center of the sun. Experiments conducted in 1919 confirmed Einstein's prediction concerning the influence of a gravitational field on the motion of light.[23,24]

*4.1 The Velocity of Light: Curvilinear Globally, Finite and Invariant Locally*

Einstein distinguished the global velocity of light in the special theory from the global velocity of light in the general theory:

> With respect to the Galileian reference body *K*, a ray of light is transmitted rectilinearly with the velocity *c* [in accord with the postulate of the special theory]. It can be easily shown that the path of the same ray is no longer a straight line when we consider it with reference to the [uniformly] accelerated chest (reference body *K'*) [relative to *K*]. From this we conclude, *that,*

# Gravitation and Electromagnetism

> *in general, rays of light are propagated curvilinearly in gravitational fields.*[25]

This is characteristic of the distinction usually made between the general and special theories concerning the motion of light. For example, in his original paper on the general theory, Einstein wrote:

> It will be seen from these reflexions that in pursuing the general theory of relativity we shall be led to a theory of gravitation, since we are able to "produce" a gravitational field merely by changing the system of co-ordinates. It will also be obvious that the principle of the constancy of the velocity of light in vacuo must be modified, since we easily recognize that the path of a ray of light with respect to $K'$ [a uniformly accelerating reference frame] must be in general curvilinear, if with respect to $K$ [a gravitation-free inertial reference frame relative to which $K'$ is uniformly accelerating] light is propagated in a straight line with a definite constant velocity.[26]

Usually, in discussions of the general theory, there is little emphasis placed on the finite and invariant velocity of light locally. Yet, Einstein did mention this last characteristic of light in the general theory in *The Meaning of Relativity* where he wrote:

> In the general theory of relativity also the velocity of light is everywhere the same, relatively to a local inertial system. This velocity is unity in our natural measure of time.[27]

Also, in his original paper on the general theory, Einstein implied in a footnote that the velocity of light locally is invariant and finite:

> We must choose the acceleration of the infinitely small ("local") system of co-ordinates so that no gravitational field occurs; this is possible for an infinitely small region. Let $X_1$, $X_2$, $X_3$, be the co-ordinates of space, and $X_4$, the appertaining co-ordinate of time measured in the appropriate unit.[*] If a rigid rod is imagined to be given as the unit measure, the co-ordinates, with a given orientation of the system of co-ordinates, have a direct physical meaning in the sense of the special theory of relativity.
> [*] The unit of time is to be chosen so that the velocity of light in vacuo as measured in the "local" system of co-ordinates is to be

equal to unity.[28]

5.   *Light and Gravitation*

If the spacetime continuum is locally Lorentzian, and the velocity of light is therefore locally invariant and finite, then the locally invariant and finite velocity of light provides a connection between the fundamental forces of gravitation and electromagnetism. The invariant and finite velocity of light is fundamental to the special theory. The special theory is essential to the development of the spacetime continuum characteristic of the general theory. Therefore, the locally invariant and finite velocity of light is fundamental to the general theory.

Contrast this conception of the motion of light with that for the motion of light globally in the general theory. Here, the generally curvilinear motion of light is dependent on its having mass, as first indicated in relativity theory in the special theory. Light, like any other physical entity, must hold to the principle of equivalence. Thus, its curvilinear motion in an accelerating reference frame, which can be equivalently considered an inertial reference frame experiencing a gravitational field, can be interpreted as the effect of the gravitational field on the light passing through it. But these interesting considerations regarding the motion of light globally do not alter the fundamental dependence of gravitation in the general theory on the finite and invariant velocity of light locally. In fact the predictions in the general theory concerning the motion of light in a gravitational field, which have been empirically supported, depend on the velocity of light being finite and invariant locally.

From a global view, in the general theory, the velocity of light is usually not invariant in reference frames. This velocity varies as spacetime is usually a curved continuum. Nonetheless, locally, the general theory relies fundamentally on the special theory. And the special theory relies fundamentally on the finite and invariant velocity of light in inertial reference frames.

Why has the dependence of the general theory on the finite and invariant velocity of light locally not been widely acknowledged? It may well be that it is difficult to maintain simultaneously that globally light travels in curved spacetime and that locally it has a finite and invariant velocity. This, though, is the case. When the dependence of the general theory on the velocity of light locally is acknowledged, and light is not considered simply another physical entity that is affected by the general theory (as happens when light is considered

# Gravitation and Electromagnetism

globally and bent by a gravitational field), the connection between the fundamental forces of gravity and electromagnetism is evident.

## *6. Conclusion*

It has been extremely difficult to integrate gravitation with the other fundamental forces of nature. It is clear, though, that gravitation depends on a fundamental characteristic of light in the special theory, namely the finite and invariant velocity of light. It is the case that this finite and invariant velocity of light holds only in local Lorentz frames, unless of course one is concerned with flat spacetime. But *in principle*, the velocity of light is finite and invariant locally, even if on a global level in the general theory it cannot be said to generally have an invariant velocity.

The reason this dependence has not been recognized is likely due to the focus on the very interesting results in the general theory concerning the velocity of light globally. This velocity stands in such contrast to the finite and invariant velocity in the special theory that it seems a contradiction that light could have a curvilinear velocity globally and yet have a finite and invariant velocity locally. These are the circumstances, though, and the finite and invariant velocity of light locally in the general theory provides the theoretical basis for connecting the fundamental forces of gravitation as developed in the general theory and electromagnetism that is elegantly discussed in the special theory.

## Endnotes

[1] Email: dsnyder@earthlink.net

[2] Some authors, including Einstein, maintain that in the special theory, it is the velocity of light that is finite and invariant in vacuum. Other authors believe it is more accurate to write that the speed of light is finite and invariant in vacuum. I believe that velocity is the better term to use. It is true that light can move in more than a single direction, thus implying that its velocity is not invariant. But, the essence of the use of the term velocity as regards light in the special theory is that for a given ray of light in inertial reference frames in uniform translational motion relative to one another, its speed and direction in vacuum do not change.

# Gravitation and Electromagnetism